\documentclass[a4paper,12pt]{article}

\usepackage{amssymb,amsmath,mathbbol,mathrsfs}
\usepackage[usenames,dvipsnames]{color}
\usepackage{hyperref}
\usepackage{xcolor}
\usepackage{stmaryrd}
\usepackage{authblk}
\usepackage{framed}
\usepackage{empheq}
\usepackage{slashed}
\usepackage{hyphenat}
\usepackage{cite}
\usepackage{apacite}

\usepackage{marginnote}    

\usepackage[left=.8in,right=.8in,top=.8in,bottom=.8in]{geometry}                  

\usepackage{sectsty} 
\allsectionsfont{\sffamily\mdseries\upshape} 
\usepackage{tocloft}

\makeatletter
\renewenvironment{abstract}{%
    \if@twocolumn
      \section*{\abstractname}%
    \else 
      \begin{center}%
        {\bfseries\sffamily\abstractname\vspace{\z@}}
      \end{center}%
      \quotation
    \fi}
    {\if@twocolumn\else\endquotation\fi}
\makeatother

\numberwithin{equation}{section}
5

\hypersetup{
	colorlinks=true,         
	linkcolor=MidnightBlue,          
	citecolor=BrickRed,        
	urlcolor=MidnightBlue            
}

\newcommand{\tocite}[1]{[{\color{magenta}\texttt{#1}}]}
\newcommand{\be}{\begin{equation}}
\newcommand{\ee}{\end{equation}}

\newcommand{\fLie}{{\mathbb{L}}}
\newcommand{\fI}{{\mathbb{i}}}

\renewcommand{\d}{{\mathrm{d}}}
\newcommand{\D}{{\mathrm{D}}}

\newcommand{\SU}{{\mathrm{SU}}}
\newcommand{\YM}{{\text{YM}}}
\newcommand{\G}{{\mathcal{G}}}

\newcommand{\pp}{{\partial}}

\newcommand{\fG}{{\mathrm{Lie}(\G)}}

\renewcommand{\bar}{\overline}
\newcommand{\dd}{{\mathbb{d}}}

\newcommand{\lbr}{\llbracket}
\newcommand{\rbr}{\rrbracket}

\newcommand{\cint}{{\int\kern-.87em{<}}}
\newcommand{\sint}{{\int\kern-.75em{\sim}}}
\newcommand{\fint}{{\int\kern-1.00em{\int}}}

\newcommand{\bb}{\mathbb}

\newcommand{\tr}{\text{tr}}

\renewcommand{\#}{\sharp}

\let\oldmarginpar\marginpar
\renewcommand\marginpar[1]{\oldmarginpar{\color{red}\raggedright\footnotesize #1}}

\title{Noether charges: the link between empirical significance of symmetries and non-separability}
\author{Henrique Gomes \footnote{\href{mailto:gomes.ha@gmail.com}{gomes.ha@gmail.com}} \\\it University of Cambridge\\ \it Trinity College, CB2 1TQ, United Kingdom}
\begin{document}
\maketitle

\abstract{A fundamental tenet of gauge theory is that physical quantities should
be gauge-invariant. This prompts the question: can gauge symmetries have
physical significance? On one hand, the Noether theorems relate conserved
charges to symmetries, endowing the latter with physical significance, though this significance is sometimes taken as \textit{indirect}. But for
theories in spatially finite and bounded regions, the standard Noether
charges are \textit{not} gauge-invariant. I here argue that
gauge-\textit{variance} of charges is tied to the nature of the
non-locality within gauge theories. I will flesh out these links by providing a chain of (local) implications: `\textit{local conservation laws}'${\Rightarrow}$ `\textit{conserved regional  charges}' $\Leftrightarrow$  `\textit{non-separability}' ${\Leftrightarrow}$ `\textit{direct empirical significance of symmetries}'.
}

\section{Introduction}

The goal of this paper is to argue for a chain of (informal) implications for gauge theories: `\textit{local conservation laws}'$\stackrel{*}{\Rightarrow}$ `\textit{conserved regional  charges}' $\Leftrightarrow$  `\textit{non-separability}' ${\Leftrightarrow}$ `\textit{direct empirical significance of symmetries}'. To forge the necessary links, I will need two raw materials: (i) the relation between symmetries and conservation, which encompass the first three arrows, and (ii) the relation between non-separability and conserved charges, which forges the third link. The link between direct empirical significance and the non-separability will here only be referred to; it is studied in depth in \cite{Gomes_new}. 

I first delineate the relevant topics in the two following subsections as a prospectus to the parts of the paper corresponding to (i) and (ii). Then section \ref{sec:roadmap} summarizes the plan.  

\subsection{Noether theorems and gauge-invariant charges}\label{sec:Noether_intro}
The theorems proved by Emmy Noether more than a century ago have had a profound impact on both physics and  mathematics. The core idea is to relate the occurrence of symmetries of the Lagrangian to conservation laws.\footnote{A good discussion of the significance of the theorems is \cite{BradingBrown_Noether}, see also the essays in   \cite{BradingCastellani} and references therein for the relation between Noether and other topics in philosophy of physics, \cite{Noether2} for a thorough treatment within physics, covering the history of the theorem and its applications  and  \cite{Noether3} for a mathematically advanced, complete analysis of the theorems.} In recent philosophy of physics, this empirical significance of symmetries---that they yield conserved charges---has been called ``indirect'', in contrast to the \textit{directly} empirically significant symmetries, exemplified by the subsystem-boosts of Galileo's ship scenario \cite{BradingBrown, GreavesWallace, Teh_emp, Friederich2014, Gomes_new}

However, in the case of gauge theories, the Noether relations do not guarantee  gauge-invariance of the local charges.  Indeed, intuition based on the Abelian case is misleading: for in this case,  the local conservation law of electromagnetism yields a well-defined  gauge-invariant conserved charge in an entire region of spacetime by integration (by the Gauss law).\footnote{We call quantities defined in regions of spacetime `regional'.  An interior quantity is a regional quantity that does not depend on boundary data.  `Regional charges' are ones obtained from integration of densities on the given regional volume and depend solely on  matter degrees of freedom, with minimal coupling. Local conservation laws are valid at each spacetime point, but may \text{not}  yield \textit{regional} conservation laws. \label{ftnt:quasi}}

Here I will use the covariant symplectic formalism---which  does not require a foliation of spacetime into equal-time surfaces---to show that in the non-Abelian case the standard Noether charge, although conserved, is  not gauge-invariant, and cannot, in general, be written solely in terms of the matter degrees of freedom. That is, beyond the lack of gauge-invariance, the generic conserved charges guaranteed by Noether's theorems will involve more than just the standard matter degrees of freedom: they will include extra contributions from the gauge field  (from the potential in the non-Abelian Yang-Mills case and from the Christoffel symbols in the gravitational case). 

I find it opportune to say a few more words on this parallel with general relativity (another nonlinear theory). There, the analogous (naive) Noether charge  is known as the Landau-Lifschitz pseudo-tensor;  this tensor is indeed  conserved, but it is coordinate-dependent and depends on the values of the Christoffel symbols in the bulk of the region.  In general relativity, one can obtain coordinate-independent conserved charges that are independent of the gravitational forces, but one can do so only in very specific models of the theory (see de Haro's contribution to this volume). Namely,  when the background metric is `uniform': that is, when the metric has associated Killing vector fields in the directions of uniformity, for each one of these Killing directions one obtains  conserved charges written as local integrals of solely the matter part of the energy-momentum tensor (minimally coupled). The same occurs for non-Abelian Yang-Mills fields: for each gauge transformation  leaving the gauge-field configuration invariant over the entire region---just as e.g. Poincar\'e transformations leave the Minkowski \textit{metric} invariant---there corresponds a conserved charge.\footnote{The notion is itself gauge-invariant: again resorting to the perhaps more familiar gravitational case, if a metric $g_{\mu\nu}$ has a certain Killing field $X^\mu$, under the action of a diffeomorphism the transformed metric will have the appropriately transformed (under the adjoint action of the diffeomorphism) vector field as its own Killing field. The same applies to non-Abelian Yang-Mills gauge-fields and their stabilizing directions. } This analysis thereby  constrains the existence of regional  gauge-invariant non-Abelian charges to very specific circumstances, as it does for general relativity.\footnote{Again following the parallel with the (perhaps more widely known) analogous  general relativity properties, gauge-invariant charges---i.e. those obtained through volume integrals of functions of minimally coupled matter degrees of freedom---are more rigidly constrained regionally than in the asymptotic regime.  In this paper I will not treat the asymptotic regime, save to say that the analysis remains completely valid, resulting  in `extra' conserved charges that do not require uniformity over the entire spacetime: they only require asymptotic uniformity \cite{ReggeTeitelboim1974} (see \cite{RielloSoft} for the corresponding statements in the formalism employed in this paper). These `extra charges' however appear only in the singular asymptotic limit, as a result of the effectively infinite separation between spatial points \cite{RielloSoft}. } 

Once we have established the existence of these charges, we will have forged the links `\textit{local gauge symmetries}'$\Rightarrow$ `\textit{local conservation laws}'$\stackrel{*}{\Leftrightarrow}$ `\textit{conserved regional  charges}', where the asterisk signals the special conditions that allow for conserved regional charges in the sense above.

The subtle point about the gauge-invariance of charges is important in establishing the remaining links between  conserved regional charges, non-separability, and Direct Empirical Significance (DES).

\subsection{Non-separability and direct empirical significance}\label{sec:DES_intro}

Gauge theory has often been described as a theory that exhibits excess structure; or, in John Earman’s words, “descriptive fluff” \cite{Earman2004}.
But when is “fluff” merely decorative and when does it perform a function?

 I have elsewhere argued---broadly in line with standard physics folklore---that the “descriptive fluff” associated to gauge theory is a sign of a particular type of non-locality, one that does not introduce any superluminal signalling. I have further argued that local gauge theories such as Yang-Mills are \textit{characterized} by this non-locality, which shows up for any choice of variables. 

Even in standard electromagnetism, employing  gauge-invariant electric and magnetic field variables,  differential constraints on the initial data prohibit an arbitrary specification of the electric field at each point. Naively, knowing the value of the electric flux on a sphere very far away can uniquely fix the total amount of charge contained in a much smaller region, \textit{at any instant}, even if the matter fields are unsupported at that far away sphere.  More mathematically, solving the constraint differential equations at a given point requires integration, or the inversion of differential operators, which in their turn require conditions on some  surrounding boundary. This is in contrast to the initial value problem for other  dynamical systems such  the  Klein-Gordon scalar field, which is unconstrained.

Indeed, the lesson of the Aharonov-Bohm experimental confirmation is that observables of the theory are given by  non-local equivalence classes.\footnote{That is, if we assume the local properties of the electromagnetic field are carried by the curvature, $F=\d A$ for $\d$ the exterior derivative and $A$ the gauge-potential 1-form, then, assuming the equivalence class of gauge potentials is given by $A\sim A+\d f$, for $f$ some scalar function, there can be an $A'$ such that $A'\not\sim A$, and yet $F=F'$, so that they are physically distinct but not locally distinguishable. Namely, whenever there exist 1-forms $\lambda$ such that $\d\lambda=0$ but $\lambda\neq \d f $ for any $f$. Such 1-forms exist in 1-1 correspondence with the first cohomology space of $\mathsf{Ker}(\d)/\mathsf{Im}(\d)$, and is thus a topological invariant, and thus non-local. Although we use quantum matter to detect such differences, this is only accidental; the fact is that the gauge-potential theory carries different physical quantities than the field strength theory.}  As emphasized by Earman in \cite{Earman2019}, a certain type of non-locality of electromagnetic phenomena \textit{is} the lesson of the experiment.  In Earman's words:
 \begin{quote}
 Thus, despite the fact
that non-simple connectedness (of the electron configuration space) is essential, by
definition, to the AB effect, it is not essential to some of the key issues to which it
served to call attention. The philosophical literature seems incapable of absorbing
this fact, as if it were under the thrall of the patently invalid inference that goes: ‘The
AB effect uses non-simple connectedness; the AB effect reveals a [certain] kind of nonlocality; ergo the nonlocality derives from non-simple connectedness.’ (p.196)
 \end{quote}  \cite{Belot1998} packs a similar message: ``Until the discovery
of the Aharonov-Bohm effect, we misunderstood what electromagnetism was
telling us about our world'' (p. 532).
 
 Thus the question is not \textit{whether} gauge theory is non-local, but  \textit{how} non-local it is. 
 In trying to answer this, using holonomies as a gauge-invariant basis of vacuum electromagnetism,  Myrwold argued, in \cite{Myrvold2010}, that when non-simply connected manifolds are decomposed into simply-connected pieces, there is information in the  physical states of the whole that is not exhausted by the information contained in the physical states of the pieces. 

Similarly---but more generally---here non-separability will be characterized  as a failure of global supervenience on regions \cite{Gomes_new}. This failure occurs when the physical states of the regions fail to determine the state of the whole, and there remains a physical variety of universe states compatible with the regional physical states.

 In \cite{Gomes_new} it was shown that such a non-separability, or failure of global supervenience, is tantamount to the direct empirical significance of symmetries. That is, certain rigid gauge-symmetries, when applied to subsystems, may yield physically distinct models of the theory. These subsystem symmetries are the generators of the residual physical variety that foils global supervenience on regions (more on this below). 

And although Myrvold's argument suffices to show non-separability for electromagnetism\footnote{The argument was applied in vacuum, where it can only detect non-separability for non-simply connected universes. But adding charges, one obtains non-separability even for simply connected regions.}---where holonomies have simple composition properties and form a basis for gauge-invariant functions---it does not suffice for non-Abelian theories. To analise Non-Abelian theories, we require new tools.

For to characterize how states of the universe supervene on the regional states, we must first be able to characterize the physical content of each region independently of the other. 
 To answer these questions, my main proposal will be that, given any bounded region, one can split the degrees of freedom  into ones that require extra data at the boundary for their definition from those that do not require further information. This distinction makes it easier to characterize `non-separability' and also non-locality in gauge theories. In the Euclidean case, the split corresponds to a decomposition of the degrees of freedom of the gauge potential  into `pure-gauge' degrees of freedom and gauge-fixed, physical degrees of freedom. In the Lorentzian case, this split maps onto  a Coulombic component for the electric field---which carries the boundary information---and the quasi-local radiative components of the electric field, which do not require boundary data but carry dynamical information. (this split was performed explicitly, also for non-Abelian Yang-Mills theories, in \cite{GomesRiello_new}; see appendix \ref{sec:gf_Lor}). For simplicitly, we  adopt a unified nomenclature and label both the Lorentzian radiative and the Euclidean gauge-fixed modes as radiative, and the Euclidean pure-gauge and Lorentzian Coulombic as Coulombic.

Once we have this decomposition at hand, we can investigate non-separability through a  gluing/reconstruction theorem. The idea is that a manifold $M$ is divided into a patchwork of regions, mutually exlusive, but jointly exhaustive.\footnote{E.g. in the simple case consisting of only two regions, this would correspond to $R_+, R_-\subset M$ such that $R_+\cap R_-=\pp R$, $R_+\cup R_-=M$ and $\pp M=\emptyset$.\label{ftnt:regions}}  Although in the Lorentzian case (but not in the Euclidean) the Coulombic information is essential for specifying the complete state of a single region,  this data becomes superfluous once one has  the (interior) radiative degrees of freedom of all the regions.  In other words, in both the Lorentzian and the Euclidean signatures,   knowledge of the pure radiative  field content of each and every region can suffice to reconstruct the complete state of the field over the entire manifold; in this case knowledge of the regional Coulombic degrees of freedom adds no relevant information. In this sense, from the global perspective, these Coulombic degrees of freedom are indeed `descriptive fluff'. From the perspective of an individual region however, we must retain these degrees of freedom for a full description. 

But there is a caveat to the claims of reconstruction above: the reconstruction goes through without the aid of Coulombic data only for simply-connected, compact manifolds, and  if \textit{one of the  two} conditions is met: (a) the  regional field configurations have no isometries, or (b) no charged matter is present anywhere.  

 When the fields have regional isometries \textit{and} charged matter is present (i.e. when neither (a) nor (b) obtains), or when the underlying manifold is not simply-connected, the radiative information of the two (simply-connected) regions does \textit{not} suffice to determine the entire global state. That is: there is gauge-invariant information in the global state which is \textit{not} contained in the union of the  regional,  \textit{strictly interior}, information.  Assuming the underlying global manifold is simply-connected, the remaining global, gauge-invariant information---that which is \textit{not} determined by the strictly interior content of the subsystems---is strictly \textit{relational} and can be mapped bijectively to (subgroups of) the charge group of the theory (e.g. U(1), or SU(N), i.e. a global Lie group). 
 
 This occurrence of subgroups of the charge group arise therefore as a physical variety of global states corresponding to the same strictly interior (i.e. radiative) states.  In \cite{GomesRiello_new, Gomes_new}, such a variety was taken to characterize a \textit{direct} empirical significance of symmetries (DES), for the variety can be construed as the result of an action of a (subgroup of the) charge group on a subsystem. Since this action takes one global state to a physically different  global state, the empirical significance of these subsystem symmetries is labeled `direct'. 
 But, most relevant for this paper, this underdetermination, or holism, occurs precisely when non-trivial gauge-\textit{invariant}, conserved matter charges also exist. In this way,  the \textit{indirect} significance of gauge symmetry associated with the Noether theorems  becomes intimately connected to the \textit{direct} significance of gauge symmetry. In this way, we forge the remaining links between `\textit{local conservation laws}'$\stackrel{*}{\Rightarrow}$`\textit{conserved regional charges}' $\Leftrightarrow$  `\textit{Non-separability}' ${\Leftrightarrow}$ `\textit{direct empirical significance of symmetries}'. 
 

 In this chain, we restrict ourselves to spacetimes that are contractible and to standard gauge theories (including general relativity, cf. De Haro's contribution in this volume). The asterisk applies to the  non-Abelian case, since local conservation laws  imply a regionally conserved matter charge only for special models of the theory.  

\subsection{Road map}\label{sec:roadmap}

The first link---that local gauge symmetries lead to local conservation laws---is the subject of section \ref{sec:Noether}.  This link of course lies within the purview of Noether's theorems; I  will summarize a version of these theorems and illustrate it with one Abelian and one non-Abelian theory. 

Through these examples, we will see that the standard charges  are only gauge-invariant for the Abelian theory; they are gauge-\textit{variant} in the non-Abelian theory. Equivalently, the charge that is conserved will depend on both the matter fields and the gauge potentials. When matter and gauge-field contributions to the Lagrangian 

I will then argue that a  gauge-invariant regional notion of charge exists (cf. footnote \ref{ftnt:quasi}), but it  is non-trivial only in the presence of Killing fields (or regional isometries).

 In section \ref{sec:regions}, I will summarize a split of regional degrees of freedom into ``radiative'' and ``Coulombic''. The radiative degrees of freedom are regionally  interior and non-local: that is, they are determinable solely from the regional content of the fields, without recourse to boundary data, but they are still non-locally defined within the region. The Coulombic degrees of freedom carry all the remaining information from outside a given region that is necessary to completely determine the state within the the region. 
 Therefore,  Coulombic degrees of freedom are  `more non-local' than the radiative degrees of freedom. With knowledge of all the radiative degrees of freedom of all the regions, it is possible to reconstruct the entire physical state (including Coulombic components). 
 
   In the conclusions (section \ref{sec:conclusions}), I put the results of the previous two sections together in my chain of implications. Local gauge symmetries lead to local conservation laws, which lead to a type of non-separability, which lead both to the gauge-invariant regional charges of \ref{sec:Noether} and to what I have elsewhere interpreted to be the `direct empirical significance of symmetries' (see section \ref{sec:DES_intro} and \cite{Gomes_new}). 

In appendix \ref{sec:covariant_symp} I will very briefly introduce the covariant symplectic formalism \cite{WittenCrnkovic, Lee:1990nz}. The formalism is extremely convenient for the treatment of conservation laws and  provides a clear path between Lagrangian/covariant concepts and Hamiltonian/canonical ones.


\section{Abelian and non-Abelian Noether regional charges}\label{sec:Noether}

In this section, I will give an abridged version of Noether's theorems that is sufficent for my purposes. To do that, I require the rudimentary notions of the covariant symplectic formalism, developed in \cite{WittenCrnkovic, Lee:1990nz}. The purpose of this section is to establish a useful, minimal version of Noether's theorem using that formalism. It can be skipped by those familiar with Noether's theorems. 

First, for simplicity, suppose we are given the contractible $d$-dimensional spacetime manifold $M$ and the Lagrangian configuration space of fields on $M$,\footnote{Also known as the \textit{field}-space, as there is no 3+1 decomposition, usually understood to be involved in the use of the term `configuration'. } $\Phi:=\{\varphi^I\in C^\infty(M, W)\}$, for $W$ some vector space (to which the superscript $I$ refers). These fields are not yet required to solve any equation of motion. In cases of interest, $\Phi$  has the structure of a (infinite-dimensional) manifold, which we can use to formulate, among other things, its tangent bundle, $\mathrm{T}\Phi$. An element of $ \mathrm{T}_\varphi\Phi$ is the tangent to a one-parameter family of configurations, $\frac{d}{d\lambda}_{|\lambda=0}\varphi^I(x, \lambda)$. By standard convention, a general such vector based at $\varphi$ is called $\delta\varphi$. In terms of a field-space basis, $\varphi^I(x)$, we can also decompose such a vector field: 
$$ \delta\varphi=\int \d x\,  \delta\varphi^I(x)\frac{\delta}{\delta \varphi^I(x)}\equiv \int  \delta\varphi^I\frac{\delta}{\delta \varphi^I},
$$
which is understood as a derivation on function(als) on $\Phi$ and henceforth we omit the spacetime dependence of the integral and of the component fields.\footnote{The components $\delta\varphi^I(x)$ are  themselves distributional, e.g. can be of the form $\kappa^I(x) \delta(x,y)$ for a smooth $\kappa^I(x)$.  }

Since we are heuristically taking $\Phi$ to be a bona-fide differentiable manifold, we can build up exterior calculus on it as well. For that, we call the field-space exterior derivative by the double-struck $\dd$, and we can extend the double-struck notation to refer to more general geometric objects in $\Phi$, such as the Lie derivative and interior contraction, $\mathbb{L}_{{\delta \varphi}}, \fI_{{\delta \varphi}}$. These should all be taken to represent field-space, e.g. variational, operations. So, for example, $\bb L_{{\delta \varphi}} F=\fI_{{\delta \varphi}}\dd F=\dd F (\delta\varphi)=\delta\varphi[F]$ for a scalar field-space functional $F:\Phi\rightarrow \bb R$ (in some differentiability class, which we need not discuss).

\subsection{Gauge transformations}
Now we move on to gauge theory as it is usually formulated. We are given a Lie group $G$ (which we will  explicitly set either to $U (1)$ or $\SU (N)$) and its Lie-algebra $\mathfrak{g}$. We have an extension of $G$ to the group of gauge-transformations,  which we denote by  $\G=C^\infty(M,G)$, whose group operation is just that of $G$ pointwise on $M$, i.e. for $g,g'\in \G$, $gg'(x)=g(x)g'(x)$,  and thus we can  define the group of infinitesimal gauge transformations, $\fG= C^\infty(M,\mathfrak{g})$, as usual. 

 We also  assume $G$ has some action (a representation) on the vector space of local field-values of $\varphi$, i.e. $V$, and thus we again pointwise define the action $g\cdot \varphi(x)=g(x)\cdot\varphi(x)\in \Phi$. That is: we define pointwise the natural action of the group of gauge transformations on field space $\G\times\Phi\rightarrow \Phi$. 
Thus field-space is foliated by gauge-orbits: namely, integral curves of field-space vectors $\delta_\xi \varphi$ for $\xi\in C^\infty(M,\mathfrak{g})$, such that $\delta_\xi\varphi:=\frac{d}{d\lambda}_{|\lambda=0}(\exp(\lambda\xi)\cdot\varphi)\in\mathrm{T}_\varphi\Phi$.
Again, if the Lagrangian in \eqref{eq:funct_L} is invariant in the direction of $\delta_\xi \varphi$, and  the Euler-Lagrange equations hold, we have:
\be
\dd\mathcal{L}(\delta_\xi \varphi)= E_I\delta_\xi\varphi^I+\nabla_\mu\theta^\mu_I(\varphi)( \delta_\xi\varphi^I)\approx\nabla_\mu\theta^\mu_I(\varphi)( \delta_\xi\varphi^I)\approx0.
\ee
Following \eqref{Noether}, using the coordinate-free notation of $\theta$ as a $d-1$ form, we have the regional conservation law: 
$$\d (\theta_I(\varphi)( \delta_\xi\varphi^I))\approx0.$$

\subsection{Noether and conserved currents}\label{sec:Noether_cons}
First, we fix our fields and the action of the gauge group. Let $\psi$ be some complex-valued scalar field,\footnote{For definiteness, we could consider Dirac fermions valued in the fundamental representation $V$ of the gauge group $G=\SU(N)$,
$
\psi^{B,b}\in\mathcal C^\infty(M, \bb C^4\otimes V)
$. To simplify notation, we will omit the subscripts $(B,b)$. 
} and the gauge potential be a Lie-algebra valued 1-form, $A\in\Lambda^1(M, \mathfrak{g})$. We will have the indices $I$ refer to the vector space $\mathfrak{g}$ (since it is also the value-space of one of the fields) and omit reference to the value space of the field $\psi$.

We define an action of $\xi\in\fG$ on these variables as:
\begin{align}
\delta_\xi A^I&:=\D \xi^I:=\pp\xi^I +[A,\xi]^I\label{eq:A_transf}\\
\delta_\xi \psi &:=-\xi\psi \label{eq:psi_g}
\end{align}
where the square brackets delimit the $\mathfrak{g}$-commutator, within which we omit Lie indices.

The Yang-Mills Lagrangian, in form language,  can be written as\footnote{Again, here $\ast$ is the Hodge-star; so $\ast \d x^\mu$ is an (d-1)-form, completed to a top d-form by the wedge product ($\wedge$) with the one-form $\D\psi$. Formulas in form language are given with the correct signs in even-dimensional spacetimes. I prioritized uncluttered formulas over complete generality. }
\be\label{eq:YM_Lag}
\mathcal{L}_\YM = -\tfrac{1}{2e^2} (F^I \wedge  \ast F_I) +  (i \bar\psi \gamma_\mu \D\psi) \wedge \ast \d x^\mu 
\ee
where $F = \d A + \tfrac12[A,A]$ and $\ast$ is the Hodge dual, $\gamma^\mu$ are Dirac's gamma-matrices, and $\D_\mu \psi = \pp_\mu \psi + A_\mu^I \tau_I \psi$,  with $\tau^I$ a basis for $\mathfrak{g}$, and $e$ is the Yang--Mills coupling constant. Also, $\bar\psi = \psi^\dagger \gamma^0$, with the understanding that $\psi$ and $\psi^\dagger$ have to be considered as two independent (complex) variables. The Lie algebra index $I$ is lowered with the Kronecker delta.

Now, we compute the field-space differential of the Lagrangian,
\begin{align}
\dd \mathcal{L}_\YM = & \big( \dd A^I \wedge (- e^{-2}\D \ast F_I - \ast J_I)\big) + \dd \bar\psi \big(i \gamma_\mu \D \psi \wedge \ast \d x^\mu\big) + \big(-i \D \bar\psi \gamma_\mu \wedge \ast \d x^\mu \big) \dd \psi \notag\\
&  + \d \big(  -e^{-2}\dd A^I\wedge \ast F_I +i \bar\psi \gamma_\mu \dd \psi (\ast \d x^\mu) \big);\label{YM_eom}
\end{align}
from which the equations of motion are easily extracted: 
\be\label{eq:eom}
{\rm EL}^I_A = - e^{-2} \D \ast F^I - \ast J^I,
\qquad 
{\rm EL}_{\bar \psi} = i \gamma^\mu \D_\mu \psi ,
\qquad 
{\rm EL}_{\psi} = ({\rm EL}_{\bar \psi})^\dagger \gamma^0.
\ee Here the matter, i.e. the fermions $\psi$, carry a $\mathfrak{g}$-current density:
\be
J^\mu = (\rho, J^i)
\quad\text{with}\quad
J^\mu_I = \bar \psi \gamma^\mu \tau_I \psi =\frac{\pp \mathcal{L}}{\pp \D_\mu\psi}\tau_I\psi\label{eq:J}
\ee
This is the current whose conservation we would like to enforce, but it is not the current whose conservation falls out of Noether's theorem.  

From \eqref{YM_eom} (and \eqref{eq:funct_L} in the appendix), one finds that the symplectic potential is given by:
\be\label{eq:YM_theta}
\theta_\YM =  -e^{-2}\dd A^I \wedge  \ast F_I + i \bar\psi \gamma_\mu \dd \psi \ast \d x^\mu.
\ee 

The covariant Noether current density $j_\xi$ is given by
\begin{align}\label{eq:conserv_Noether}
j_\xi = \fI_{\xi^\#} \theta_\YM 
& = - e^{-2}\D \xi^I \wedge \ast F_I - i \bar\psi \gamma_\mu  \xi \psi \ast \d x^\mu \notag\\
& = (e^{-2}\D \ast F +\ast J)_I\xi^I - \d( \ast F_I \xi^I)\notag\\
&\approx - e^{-2}\d (\ast F_I \xi^I).
\end{align}
\subsubsection{Local and regional, Abelian and non-Abelian conservation laws}\label{sec:Abel}

 The conservation equation $\d j_\xi \approx 0$ of the standard Noether current trivially follows from \eqref{eq:conserv_Noether}, but it is not a gauge-covariant conservation law. That is, it obeys: $\pp_\mu j^\mu_I=0$ but \textit{not} $\D_\mu j^\mu_I=0$
 
 To obtain the relation of the Noether to the matter current, $J^\mu_I$, we insert a constant $\xi^I(x)=\xi^I_o$, i.e. such that $\pp_\mu\xi^I=0$ in the first line of \eqref{eq:conserv_Noether}.  Then the $\xi$-independent relation between the currents is:
 \be\label{eq:Noether_j}
 j^\mu_I=-[A_\nu, F^{\mu\nu}]_I+J^\mu_I
 \ee
 
 To arrive at a regional conservation law, we must integrate this current over a region. 
 But here we see that, upon integration,  the conservation for $j^\mu$ would involve not only the values of the matter fields and current in the bulk of a region, $J^\mu_I$, but \textit{also} the values of the gauge potential and curvature there! This failure to decouple the conservation of fermionic matter from bosonic fields implies that we cannot easily  relate the bulk integral of quantities involving matter fields with the boundary fluxes of quantities involving the  forces. Thus, the relation \eqref{eq:Noether_j} so far has no implications for non-locality: we must know the values of the gauge-fields everywhere, not only at a distant boundary, to infer the total amount of matter charge contained in a region.


 To remedy this situation, we first obtain a current that obeys a local covariant conservation law. Of course, $J$ fills this role. Therefore, by applying the Bianchi identities to the equations of motion for $A$:\footnote{While it is standard to just define $J_\mu^I$  as the source of the Yang-Mills equations of motion  \eqref{eq:eom}, and derive the  covariant laws  from  $\D\ast \D \ast F=0$, which vanishes by its algebraic symmetry properties, we can recover a related result. First, if the action functional is gauge-invariant, a variation in the direction of $\delta A=\delta_\xi A$ will not change it.  
 With minimal coupling, assuming the Lagrangian has a kinetic term for the field $A$ which does \textit{not} depend on the matter fields, if we defined $\tilde J_\mu^I:=\frac{\pp \mathcal{L}^{\text{\tiny{matter}}}}{\pp A_\mu^I}$ we would obtain from  \eqref{eq:eom}:
$$ \dd S_\YM(\delta_\xi A)= \int \mathrm{Tr}(({\rm EL}^{\text{\tiny{gauge}}}_A+{\rm EL}^{\text{\tiny{matter}}}_A) \D\xi)\,{=} -\int  \mathrm{Tr}(\ast\D\ast({\rm EL}^{\text{\tiny{gauge}}}_A+ \tilde J)\xi)=0.
$$
 The anti-symmetry of the spacetime indices of the curvature tensor guarantee that $\ast\D\ast {\rm EL}^{\text{\tiny{gauge}}}_A=0$ and therefore that $\ast\D\ast\tilde J=0$. But we must use the equations of motion if we want to say that $\tilde J=J$, that is: $J_\mu^I:=\frac{\pp \mathcal{L}^{\text{\tiny{matter}}}}{\pp \D_\mu\psi}\tau_I\psi\approx\frac{\pp \mathcal{L}^{\text{\tiny{matter}}}}{\pp A_\mu^I}$. The same type of relation arises in general relativity, but if one actually defines the matter energy momentum tensor as the functional derivative of the matter Lagrangian by the metric, then, using the Bianchi identity, no equations of motion need to be used. In this way, it becomes clear that merely employing gauge variables, we can recover the necessary local conservation laws from variational principles.\label{ftnt:current}}
 \be\label{eq:Noether_covJ}\D_\mu J^\mu_I\approx0
 \ee
  In the Abelian case, there is no difference between $j$ and $J$, of course, so the conservation equation is automatically covariant. 
  But now, at least at first sight, we have lost the possibility of integrating this current to obtain a regional conservation law, and so no non-locality is apparent.
 
 Crucially, to determine regional conservation laws, we need not the spatially constant gauge-transformations, generated by $\pp_\mu\xi^I=0$, but those generated by the \textit{covariantly constant} $\D_\mu \xi^I=0$.

In other words,  we can integrate $\D_\mu J^\mu_I$ against any Lie-algebra valued scalar $\xi\in C^\infty(M, \mathfrak{g})$ on the bounded region $R$, obtaining: 
\be \label{eq:quasi_charge}
\int_N  (\D_\mu J^\mu_I\xi^I)=-\oint_{\pp N}n_\mu \,J^\mu_I \xi^I+\int_N J^\mu_I \D_\mu \xi^I=0
\ee
But this is  a bona-fide regional conservation law if $\xi^I$'s are such that $\D_\mu \xi^I=0$, for otherwise the integral will not generically reduce to a boundary flux.

This establishes some caveats to our usual understanding of the Noether conserved currents. Such caveats are acknowledged,\footnote{Brading and Brown describe this finding as \cite{BradingBrown_Noether}:``we cannot follow the procedure used for global gauge
symmetries in the case of local gauge symmetries to form gauge-independent currents. A current that is dependent on the gauge transformation parameter is not satisfactory ---in particular,
such gauge-dependent quantities are not observable.'' } but their more radical consequences  are usually left unsaid: (i) ultimately, the very concept of a non-Abelian regionally conserved charge  only makes sense over a uniform background (as the notion of regionally conserved energy-momentum in general relativity) and (ii) Noether charges are not invariant in the presence of  boundaries to the Cauchy surface.

  There is a useful parallel to be drawn here  with general relativity. 
In general relativity, one can use the same procedure as above to find $j_\mu^I\sim \tau_{\mu\nu}$, the so-called Landau-Lifschitz pseudo-tensor  (see e.g.: \cite[Ch. 15.3]{WeinbergQFT2} and \cite[Ch. 11]{LandauLifschitzVol2}).  But what we want is a covariant local conservation law, such as $\nabla_\mu T^{\mu\nu}=0$. Indeed,  the energy-momentum tensor  $ T_{\mu\nu}$ defined in similar fashion to $J^I_\mu$ in \eqref{eq:J} does satisfy this covariant local conservation law. But this only  straightforwardly yields   regional conservation laws   if Killing vector fields are available,\footnote{Here I am focusing on notions of charge which arise from integration over volume. Common definitions of `quasi-local' charges---such as the ones studied by Szabados, e.g. Brown-York---fall outside of this definition. Although there are many suggestions for notions of conserved `quasi-local' energy-momentum, it is fair to say no one notion is as of yet unproblematic. As Szabados remarks in his review \cite[p. 9]{Szabados2004}: ``Contrary to the high expectations of the eighties, finding an appropriate quasi-local
notion of energy-momentum has proven to be surprisingly difficult. Nowadays, the state of the art
is typically postmodern: Although there are several promising and useful suggestions, we have not
only no ultimate, generally accepted expression for the [quasi-local] energy-momentum and especially for the
angular momentum, but there is no consensus in the relativity community  [...]
 on the list of the criteria of reasonableness of such expressions. The various suggestions
are based on different philosophies, approaches and give different results in the same situation.''   } i.e. if there is a $\xi_\mu$ such that $\nabla_{(\mu}\xi_{\nu)}=0$; for then we can write a smeared equation in a bounded spacetime region $N\subset M$:
\be \int_{N} \nabla_\mu T^{\mu\nu}\xi_\nu=\oint_{\pp N} n_\mu T^{\mu\nu}\xi_\nu=0\label{eq:spt_cons}
\ee
where $n_j$ is the normal to $\pp N$; this yields a regional conservation law in the same way as above. 

As in general relativity, in the non-Abelian Yang-Mills case generic configurations do not admit any such Killing fields, or directions of uniformity. Therefore, true regional conservation laws emerge only in the presence of matter charges \textit{and} directions of uniformity in the background field. 

In the chain of implications, we can formalize the asterisk as the following condition:
$$*\quad:\quad\text{over field configurations $\tilde A$ such that there exist $\tilde\xi_I$ with $\tilde\D\tilde\xi=0$}
$$
(mutatis mutandis for general relativity).
Thus we have secured the link `local conservation laws'$\stackrel{*}{\Rightarrow}$ `conserved regional charges', within the special models signalled by the asterisk. Although the link would also apply more generally for any field theory employing gauge potentials and matter fields that transform as in \eqref{eq:A_transf} and \eqref{eq:psi_g} (on-shell of the matter equations of motion), we will not show this here. 

As mentioned in section \ref{sec:DES_intro}, according to \cite{Gomes_new}  regional conservation laws are related to the type of failure of global supervenience on regions required for the existence of `symmetries with direct empirical significance' (a thesis first articulated in \cite[Sec 4.4]{GomesRiello_new}). In this respect, the important point  to secure this relation is \textit{not} that the charges $J$ be of one specific form or another; the important point is that the conserved regional charges be associated to the Killing directions $\xi^I$, which are rigid. 
In the next section, we will examine just how this fact is related to non-separability and therefore with the direct empirical significance of symmetries. 



\section{Non-separability}\label{sec:regions}

Here we will show that  for simply-connected manifolds conserved regional charges occur just in case there is a certain level of non-separability in the model. Following Myrvold, we will first investigate this matter in electromagnetism using holonomies, but adding charged scalar fields. This will make it clear, at least in electromagnetism, that there is a residual variety in  reconstructing the global state from the states of the regions  if and only if charges are present within the regions.  

But holonomies are not sufficient to settle the question in the non-Abelian case, and thus we will resort to a gauge-fixing, in section \ref{sec:gf_Euc}. Once we have gauge-fixed the regional content of the field, we will use a gluing theorem to reconstruct a gauge-fixed state of the whole from that of the regions. It will be the case once more that, in the presence of regional conserved charges, one cannot reconstruct the state of the whole from the state of the parts, thus establishing the link between `regional conserved charges'$\Rightarrow$`non-separability'. 

Section \ref{sec:gf_Euc} applies in the Euclidean signature regime. But in such a regime, matters of locality are orthogonal to matters of separability; there are in fact no matters of locality, since there is no causal propagation.  
But, as separability suffices to forge the link of our chain, we leave an investigation of the Lorentzian models to appendix \ref{sec:gf_Lor}. There, we will locate the source of non-locality of gauge theory in the Gauss constraint. Thus to be able to discuss the state of one region independently of the state of another---a necessary independence to analyse separability---we will split the variables into a Coulombic part, that solves the Gauss constraint as in the electrostatic case, and a radiative, dynamical part. The Coulombic part is the one that relates the charge content in the bulk of the region to boundary information. However, it is only the radiative part that enters the reconstruction theorem, and the theorem suffers from the same degeneracy as in the Euclidean case.  

\subsection{Holonomies and non-separability}

The holonomy interpretation of electromagnetism takes as its basic elements assignments of unit complex numbers to loops in spacetime. A loop is the image of a smooth  embedding of the oriented circle,  $\gamma:S^1\rightarrow \Sigma$; the image is therefore a closed, oriented, non-intersecting curve. One can form a basis of gauge-invariant quantities for the holonomies (cf. \cite{Barrett_hol} and \cite[Ch.4.4]{Healey_book} and references therein),\footnote{Of course, any discussion of matter charges and normalization of action functionals would require $e$ and $\hbar$ to appear. However, I am not treating matter, so these questions of  choice of unit do not become paramount. As before, if needed, I set my units  to $e=\hbar=1$; as is the standard choice in quantum chromodynamics (or as in the so-called Hartree convention for atomic units).} 
\be\label{eq:hol} hol(\gamma):=\exp{(i\int_\gamma A)}.\ee

\subsubsection{The basic formalism}
Let us look at this in more detail.  By exponentiation (path-ordered in the non-Abelian case), we can assign a complex number  (matrix element in the non-Abelian case) $hol(C)$ to the oriented embedding of the unit interval: $C:[0,1]\mapsto M$. This makes it easier to see how composition works: if the endpoint of $C_1$ coincides with the starting point of $C_2$, we define the composition $C_1\circ C_2$ as, again, a map from $[0,1]$ into $M$, which takes $[0,1/2]$ to traverse $C_1$ and $[1/2, 1]$ to traverse $C_2$.  The inverse $C^{-1}$ traces out the same curve with the opposite orientation, and therefore $C\circ C^{-1}=C(0)$.\footnote{It is rather intuitive that we don't want to consider curves that trace the same path back and forth, i.e.  \textit{thin} curves. Therefore we  define a closed curve as \textit{thin} if it is possible to shrink it down to a point while remaining within its image. Quotienting the space of curves by those that are thin, we obtain the space of \textit{hoops}, and this is the actual space considered in the treatment of holonomies.   I will not call attention to this finer point, since it follows from a rather intuitive understanding of the composition of curves.} 
Following this composition law, it is easy to see from \eqref{eq:hol} that 
\be\label{eq:loop_com} hol(C_1\circ C_2)=hol(C_1)hol(C_2),\ee with the right hand side understood as complex multiplication in the Abelian case, and as composition of linear transformations, or  multiplication of matrices, in the non-Abelian case.

For both Abelian and non-Abelian groups, given the above notion of composition, holonomies are conceived of as smooth homomorphisms from the space of loops into a suitable Lie group. One obtains a representation of these abstractly defined holonomies as
holonomies of a connection on a principal fiber bundle with that Lie group as structure group; the collection of such holonomies carries the same amount of information as the gauge-field $A$. However, only for an Abelian theory can we cash this relation out in terms of gauge-invariant functionals. That is, while \eqref{eq:hol} is gauge-invariant, the non-Abelian counterpart (with a path-ordered exponential), is not.\footnote{For non-Abelian theories the gauge-invariant counterparts of \eqref{eq:hol} are Wilson loops, see e.g. \cite{Barrett_hol}, 
$ W(\gamma):=\text{Tr}\, \mathcal{P}\exp{(i\int_\gamma A)}
$,
where one must take the trace of the (path-ordered) exponential of the gauge-potential. It is true that all the gauge-invariant content of the theory can be reconstructed from Wilson loops. But,  importantly for our purposes, it is no longer true that there is a homomorphism from the composition of loops to the composition of Wilson loops. That is, it is no longer true that the counterpart \eqref{eq:loop_com} holds,  $W(\gamma_1\circ\gamma_2)\neq W(\gamma_1)W(\gamma_2)$. This is  due solely to the presence of the trace. The general composition constraints---named after Mandelstam---come from generalizations of the Jacobi identity for Lie algebras, and depend on $N$ for SU($N$)-theories; e.g. for $N=2$, they apply to three paths and are: $W(\gamma_1)W(\gamma_2)W(\gamma_3)-\frac12(W(\gamma_1\gamma_2)W(\gamma_3)+W(\gamma_2\gamma_3)W(\gamma_1)+W(\gamma_1\gamma_3)W(\gamma_2))+\frac14(W(\gamma_1\gamma_2\gamma_3) + W(\gamma_1\gamma_3\gamma_2) = 0$.  \label{ftnt:tr_hol}}

\subsubsection{Non-separability}\label{sec:hol_charge}
As both Healey \cite[Ch. 4.4]{Healey_book} and Belot (\cite[Sec.12]{Belot2003} and \cite[Sec.3]{Belot1998}) have pointed out, even classical electromagnetism, in the holonomy interpretation, evinces a form of non-locality, which one might otherwise have thought was a hallmark of non-classical physics. But is it still the case that  the state of a region supervenes on assignments of intrinsic properties to patches of the region (where the patches may be
taken to be arbitrarily small)? This is essentially the question of \textit{separability} of the theory (see \cite[Sec.3]{Belot1998}, \cite[Sec.12]{Belot2003},  \cite[Ch.2.4]{Healey_book},  and \cite{Myrvold2010}). 
\begin{figure}[t]
		\begin{center}
			\includegraphics[width=7cm]{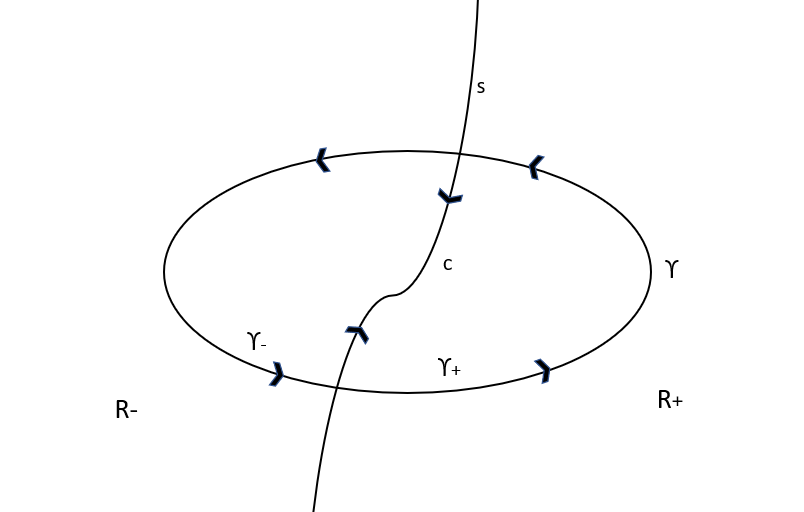}	
			\caption{ Two subregions, i.e. $R_\pm$,  with the separating surface $S$. We assume no matter fields are present on $S$. A larger loop $\gamma$ crosses both regions. But, since $\gamma_1$ and $\gamma_2$ traverse $S$ along $C$ in opposite directions, $\gamma=\gamma_1\circ\gamma_2$.
			}
			\label{fig4}
		\end{center}
\end{figure}

 We are not  interested here in cases of ``topological holism'', as related to the Aharonov-Bohm effect, which has been extensively studied. We are asking whether a vacuum, simply-connected universe still displays non-separability. For this topic, we can directly follow Myrvold's definition \cite[p.427]{Myrvold2010} (which builds on Healey's notion of Weak Separability \cite[p. 46]{Healey_book} and on Belot's notion of Synchronic
Locality \cite[p. 540]{Belot1998}):\\ \smallskip

\noindent$\bullet$ \textit{Patchy Separability for Simply-Connected Regions}. For any simply-connected spacetime region $R$, there are arbitrarily fine open coverings $\mathcal{N}=\{R_i\}$ of $R$ such that the state of $R$ supervenes on an assignment of qualitative intrinsic physical properties to elements of $\mathcal{N}$.\\

In vacuum,  it is easy to show that  \textit{Patchy Separability for Simply-Connected Regions} holds. In Figure 2, we see a loop $\gamma$ not contained in either $R_+$ or $R_-$. However, we can decompose it as $\gamma=\gamma_+\circ\gamma_-$, where each regional loop $\gamma_\pm$ does not enter the complementary region ($R_\mp$, respectively). Following \eqref{eq:loop_com}, it is then true that, since holonomies form a basis of gauge-invariant quantities,  the universal gauge-invariant content of the theory supervenes on the regional gauge-invariant content of the theory.\smallskip

It is also easy to see how \textit{Patchy Separability for Simply-Connected Regions} fails when charges are present within the regions but absent from the boundary $S$  (see in particular  \cite[Sec. 4.3.2]{GomesRiello_new}, and footnote 70 in \cite{Gomes_new}). For, in the presence of charges, we can form \textit{gauge-invariant} functions from a non-closed curve $C'$  that crosses $S$ and e.g. one positive and one negative charge, $\psi_\pm(x_\pm)$, capping off $C'$ at $x_\pm\in R_\pm$. That is, the following quantity is a gauge-invariant function:
$$Q(C', \psi_\pm)= \psi_-(x_-)hol(C')\psi_+(x_+)
$$
for $C'(0)=x_-, C'(1)=x_+$. It is easy to check from the transformation property $\psi\mapsto g\psi$, that $Q$ is gauge-invariant. But now,  applying different rigid gauge transformations $\theta_\pm$, in each region will amount to choosing a different regional representation for $\psi_\pm$, that is, $e^{i\theta_\pm}\psi_\pm$, thus resulting in a relative shift $Q\rightarrow e^{i(\theta_+-\theta-)}Q$ (which, as stated above, is gauge-invariant and therefore cannot be undone by a gauge transformation). This is precisely what is known in the literature as `` `t Hooft's beam-splitter''  (see \cite[p.110]{thooft} and and  \cite[p. 651]{BradingBrown}).

If the reader is content with the Abelian case, they may feel comfortable with skipping the next subsections. 
\subsection{Gauge-fixing regional and universal states}\label{sec:gf_Euc}

In the general non-Abelian case for bounded regions, a natural gauge-fixing suggests itself.\footnote{As we will see in appendix \ref{sec:gf_Lor}, this gauge-fixing is related to the split of electric degrees of freedom into those that solve the Gauss constraint as it is solved in the static case, and those that are radiative, or dynamical. It is also naturally selected by the ultralocal inner-product in the field space of the theory, $\Phi$. }
Given a field space vector $\delta A\in T_A\mathcal{A}$, where $\mathcal{A}=\Lambda^1(M, \mathfrak{g})$ is the  space of (kinematical) models for the gauge-potential, we define the Landau gauge-fixing as:
\be\label{eq:Landau}
\D^\mu\delta A_{\mu I}=0
\ee
In the bounded case, suppose  that our patchwork of mutually exclusive, jointly exhaustive regions is given by just $R_+, R_-$,  such that $ R_+\cap  R_-=\pp  R_\pm$, with $s^i$ the normal to $\pp R$, and  where  $ R_+\cup R_-=\Sigma$.  We also assume $\Sigma$ to be simply-connected and without boundary, $\pi_1(\Sigma)=0, \pp\Sigma=\emptyset$. Suppose moreover that there are is no charged matter  at $\pp R$. 

In that case, the gauge-fixing \eqref{eq:Landau} becomes:\footnote{See \cite[Sec.2.2.2]{DES_gf} for how the natural extension of the Landau gauge-fixing to the bounded case implies the Neumann boundary conditions used here.}
\be\label{eq:h_euc}
\begin{cases}
\D^\mu\delta A^\pm_{\mu I}=0\\
\delta A^\pm_{sI}=0
\end{cases}
\ee
where, again,  $s$ is the normal to the shared boundary of the region. We call a field-space vector that satisfies these conditions, horizontal (or radiative, in the Lorentzian case), and denote them by $h$ and $h_\pm$, in the global and regional cases, respectively. 

It is easy to show that \eqref{eq:Landau} and \eqref{eq:h_euc} are bona-fide perturbative, gauge-fixings, that are moreover covariant under a change of the perturbed configuration $A\rightarrow A+\delta_\xi A$ (and mutatis mutandi for the regional case, so that the  gauge-fixing of the perturbation does not sensitively depend on the choice of representative for the perturbed configuration. 

Two configurations $h_\pm$ are composable if and only if there exist gauge transformations, $\xi_\pm$, such that:
$$\delta A=(h_++\D_+\xi_+)\Theta_++(h_-+\D_-\xi_-)\Theta_-,$$  where $\Theta_\pm$ are the characteristic functions of the regions $R_\pm$, is a smooth global gauge configuration. Composability therefore requires the difference of the configurations at the boundary, $(h_+-h_-)_{|S}=(\D_+\xi_+-\D_-\xi_-)$, to be pure gauge. These matters are taken up in detail in \cite{GomesRiello_new, Gomes_new, DES_gf}. 

In general there will be many such $\xi_\pm$ if there is any. But to establish residual physical variety of the global state,  we must also consider the variety of global \textit{physical} states that will correspond to the composition, and therefore we must gauge-fix globally, with \eqref{eq:Landau}. In most cases, this global gauge-fixing will uniquely fix $\xi_\pm$; when it does not, there can be a residual physical variety, as we explain next. 

\subsection{Coulombic and radiative reconstruction theorem}\label{sec:gluing}

To finish this section, I will present  results on `gluing':  in reassembling the regional information of the fields into the global state of the fields, employment of the  pure-gauge and horizontal (alternatively, Coulombic and radiative, cf. appendix \ref{sec:gf_Lor}) split of  regional degrees of freedom is remarkably instructive. (To avoid having to intersperse such parentheses along the following section, the results are framed in the unifying language of `Coulombic and radiatives' modes). Again, I will only state, and not prove any of the surprising results here (see \cite[Sec. 4]{GomesRiello_new}). 

\subsubsection{The theorem, loosely stated}

 The reconstruction theorem states that in these circumstances, we can fully and uniquely reconstruct the state on the entire region, e.g.  we can recover  the full physical content of the fields,\footnote{In the non-Abelian case, we recover $E$ up to gauge transformations, but we only recover the physical content of a perturbation of the potential, $\delta A$. We can recover the full $A$, and not just perturbations, in the Abelian case. 
 This reconstruction  works either in the dynamical 3+1 context in which $M$ represents a Cauchy surface, or for the Euclidean covariant context in which $M$ is the spacetime manifold. In the Euclidean context (described in \cite{GomesHopfRiello}), we don't talk about the phase space $\mathcal{P}$, but solely about the Lagrangian field space $\Phi$, discussed in section \ref{sec:cov_form}.}  solely from  the \textit{strictly interior} degrees of freedom. Those degrees of freedom that require further boundary  information---the Coulombic---are superfluos for reconstruction.  
 
 As a schematic example, in the Lorentzian case (cf appendix \ref{sec:gf_Lor}),  we again split the electric field and the spatial gauge potential along radiative and Coulombic components. The Coulombic component of $E$ generalizes the electrostatic electric field, and is of the form $E_\text{Coul}=\D\Lambda$, for some (Lie-algebra-valued) scalar potential $\Lambda$ and $E_{\text{rad}}$ being divergence-free, as in \eqref{eq:Landau}: $\D^iE_{iI}=0$ (obeying the same Neumann boundary conditions as in \eqref{eq:h_euc}).
 
 Thus the total regional field is $E_\pm=E^\text{rad}_\pm+\D\Lambda_\pm$, with $\Lambda_\pm$ satisfying the Poisson equation $\D^2_\pm\Lambda_\pm=\rho_\pm$, with  $\rho_\pm$ related to the regional matter fields $\psi_\pm, \bar\psi_\pm$  by \eqref{eq:J},  with  boundary conditions on the Poisson equation given by $s^i\D_i\Lambda=s^iE_i\equiv f$.  It would seem the Coulombic degrees of freedom parametrized by $\Lambda_\pm$ are essential for a description of the global state. Surprisingly, we can recover the global $E$ solely as a functional of those degrees of freedom which carry no dependence on the boundary conditions (i.e. the radiative ones). 
 
 The statement of the gluing theorem therefore is that, in the Euclidean case, if the perturbed potential $A^\pm_{\mu I}$ has either no Killing directions (i.e. no stabilizers),  or no charged currents inside the regions $R_\pm$, then $h_\pm$  uniquely determines the global physical state $h$. In the Lorentzian case (for which now $R$ and $R_\pm$ are spatial regions), the statement is that if the phase space points $(A^\pm_{iI}, E_i^{I\pm})$ have no stabilizers, the radiative components $E^\text{rad}_\pm$ and the matter charges $\psi, \bar\psi$, suffice to determine all the remaining regional components, $E^\text{Coul}_\pm$, and also the global ones:  $E^\text{rad}$ and $E^\text{Coul}$.
 
  For the proof, which uses the remarkable properties of the Dirichlet-to-Neuman operators, see \cite[Sec 5]{GomesRiello_new}.

 \subsubsection{The caveats, loosely stated}
 But, as announced, if the necessary conditions do not obtain, there are ambiguities in the reconstruction. These occur only if the underlying fields   $A^{\mu I}_\pm$, or, in the Lorentzian setting, $(A^{iI}_{\pm}, E^{iI}_{\pm})$, admit  stabilizers (or Killing vectors). Let us call such a homogeneous, or uniform potential,  $\tilde A^{\mu I}_\pm$; the arising ambiguities are precisely the stabilizers of $\tilde A$,  that we label $\chi_{\tilde A}$.
 
A gauge transformation by one such stabilizer  does not affect the gauge fields (by construction, since $\D\xi=0$), but  does affect the matter fields, by the transformation law \eqref{eq:psi_g}. Therefore If these ambiguities exist,  \textit{and} there are non-trivial matter fields inside the regions, a corresponding ambiguity in the reconstruction slips in: the matter components \textit{are} shifted by an element of the stabilize. Although this rotation does not affect either regional gauge-invariant state, it has a relative effect between the regions.  This is precisely the same sort of regional rigid gauge shift we saw at the end of section \ref{sec:hol_charge}, and to the `t Hooft beam-splitter. And these conditions are precisely the ones required for the existence of `conserved regional charges'. Therefore we have established the link `conserved regional charges'$\Rightarrow$`non-separability'. 

  It is also possible at this stage to understand why in the Abelian, but not in the non-Abelian case, we always find well-defined regional gauge-invariant charges (in the presence of charged matter), thus connecting the present discussion back to section \ref{sec:Abel}. Namely,  for Abelian theories the existence of stabilizers $\chi$ impose no restrictions on $A$: for all $A$, the stabilizers are just the constant gauge transformations.

The same argument could be applied to general relativity: suppose there are matter fields that vanish at the intersection of two flat regions, but not in their interior.  Then we could gauge-fix all the non-rigid diffeomorphisms with the metric field, but that would leave out the rigid Poincar\'e transformations. Then applying distinct such transformations to each region, we would have no global Poincar\'e transformation that would be able to efface the difference between the rigid regional transformations at the boundary. Although without the presence of matter the overall state of affairs after the tranformation would be the same as before, it would be distinct if matter was present. Those are again precisely the conditions required to obtain non-trivial regional conserved charges in general relativity.

\subsubsection{The link to DES}
To establish the further link to DES, we note that the ambiguity corresponds to a global---or rigid, in the nomenclature of \cite{Gomes_new}---gauge transformation, which is necessarily a subgroup of the charge group $G$. See \cite[Sec. 5.2.4]{Gomes_new} and \cite[Sec. 4]{GomesRiello_new} for proofs and explicit examples. In the Abelian case of electromagnetism, there is always a conserved U(1) global transformation if charged matter fields are present in the regions.\footnote{We note that it is uncontroversial that DES, in many of its incarnations, e.g. \cite{BradingBrown, GreavesWallace, Teh_emp, Friederich2014}, is conditional on the existence of universal gauge-invariant quantities that are \textit{not} specified by the regional gauge-invariant content. It is thus easy to see that non-separability gets us almost all the way to DES; all that remains to be shown is that the variety is parametrized by the action of a rigid symmetry group, which we did obtain in the gluing theorem above.  The exact entailment between the two concepts---non-separability and DES---is fleshed out in full in \cite{Gomes_new}.}
 
 To summarize, to obtain a left-over variety in the reconstruction of the whole from the radiative parts, we require non-trivial stabilizers and non-zero charged matter fields. Moreover, we can explicitly  associate  this physical variety with the action of a subgroup of $G$ on a subsystem. This action can indeed be construed as endowing  `direct empirical significance' to this group of global symmetries \cite{GomesRiello_new, Gomes_new}. Lastly, the requirements for obtaining this variety also  correspond to those for obtaining non-trivially conserved regional, gauge-invariant Noether charges. 
  

\section{Conclusions}\label{sec:conclusions}

\subsection{Summary}

Local conservation laws are a widely celebrated consequence of symmetries. While it is true that in certain cases one can extract conserved \textit{regional}  quantities from integrating the Noether densities,   such charges may be problematic in other regards. For instance, the obtained regional charges may  be gauge-\textit{variant}. Or, when we think of fermionic charges as sourcing bosonic forces, it could be that we cannot isolate a regional conservation law for the matter charges. Both of these obstacles occur for non-Abelian Yang-Mills theory and general relativity, and both can be avoided at a single stroke:  with gauge-invariant conserved regional matter charges emerging from the regional volume only if the underlying configuration admits stabilizers.\footnote{In Abelian theories, any gauge field configuration admits the same stabilizers: the constant gauge transformation. This property gives rise to well-defined regional electric charges independently of the background configuration in electromagnetism. }  
This result provided the first link of the chain of implications proposed in this work. 

The second link connected such regional charges with non-separability. This link followed without ancillary conditions and applies also to simply-connected regions (in the non-simply-connected case, the vacuum configurations are also non-separable by the same methods). We showed it first for the Abelian theory using holonomies and then in the non-Abelian case using gauge-fixings. Crucially, non-separability in the simply-connected case requires a gauge transformation that is an isomorphism of the gauge field but not of the matter field. If the matter field vanishes \textit{anywhere}, such transformations cannot be gauge-fixed, and yet, if different such transformations are applied in each region, they will change the physical state of the union of the regions. Nonetheless, from the perspective of the region, the shifted configuration is physically indiscernible from the unshifted configuration. Thus, we obtained non-separability, and clearly, a connection with a subsystem symmetry that could be said to have empirical significance. 
Although \cite{Gomes_new} fleshes out this link in more detail, here we saw a (hopefully convincing) sketch of the link, listing its main arguments.   

Although the paper was focused on Yang-Mills theories, the main arguments here can be transposed without much difficulty to general relativity. In the interest of space, I have kept this parallel mostly in the background for this paper. 

Moreover, a few words should be said about the generality of these links: in which sense are they merely kinematical, and in which do they depend on the specific form of the Lagrangian? The only step in which we explicitly used the Lagrangian was in obtaining the regionally conserved charges. But, for gauge-invariant theories that have $A$ and $\psi$ as variables, with the transformation properties \eqref{eq:A_transf} and \eqref{eq:psi_g}, that link follows as well (on-shell of the matter equations of motion, cf. footnote \ref{ftnt:current}). But I have not explored the full space of theories in which these links would hold, and therefore the implications need to be seen as ``local deductions''. 

\subsection{An added link to locality}

Local symmetries have another, less celebrated consequence: they yield constraints on the initial values, thus pointing  to a mild form of non-locality in a gauge theory. As discussed in appendix \ref{sec:gf_Lor}, one cannot freely specify the initial value of the electric field here without knowing what is the far away distribution of charges  at that same instant. 

The link to non-locality does not fit as neatly into the chain of deductions explored in this paper, although it is inextricably connected to them. So, lest we tie them together, we would be left with all those loose ends; thus the inclusion of the discussion in the appendix \ref{sec:gf_Lor} (but not on the main paper) and this extended subsection here in the conclusions.

  The particular non-locality of gauge theories can be examined precisely with the aid of a  decomposition of the fields into Coulombic and radiative components. The radiative component corresponds to the strictly interior field content of a region; the Coulombic component depends  also on boundary  information. 

We can track the effect of the decomposition on the Noether charges of the theory: the emerging radiative Noether charges are identical to the gauge-invariant Noether charges discussed in the first sections. The Coulombic charges can be identified with the  gauge-variant charges; they correspond to the boundary gauge group: $\fG_{\pp R}$. 

In the absence of stabilizers,  the radiative modes are the sole carriers of the non-redundant regional information needed to piece back together a global state. The Coulombic degrees of freedom can also be reassembled by knowledge of the regional \textit{radiative} degrees of freedom and are, in that sense,  superfluous. But they become superfluous only from the perspective of the whole; given any single region, they are functionally independent from the radiative modes \cite[Sec 5]{GomesRiello_new}. This analysis is in accord with the proposal of \cite{GomesStudies}, which argues for the ineluctability of redundant, pure gauge degrees of freedom for the description of regions, but not for the description of the whole.

\subsection*{Acknowledgements}
I would like to thank an anonymous reviewer, for very insightful and constructive remarks. 




\appendix
\section*{APPENDIX}

\section{Symplectic formalism}\label{sec:covariant_symp}

\subsection{The covariant symplectic formalism abridged}\label{sec:cov_form}

A given action 
$$S[\varphi, \gamma]=\int_M\mathcal{L}(\varphi^I, \nabla_\mu\varphi^I, \nabla_{(\mu}\nabla_{\nu)}\varphi^I,  \nabla_{(\mu_1}\cdots\nabla_{\mu_n)}\varphi^I,  \gamma)$$
for the Lagrangian density $\mathcal{L}$ dependent on the  field $\varphi^I$, its derivatives up to n-th order,\footnote{We can use symmetrized derivatives because we may always rid ourselves of the anti-symmetrized ones in favor of the appearance of curvature, which here would be seen as part of the background structure. We denote spacetime indices by Greek letters, $\mu, \nu, \cdots$.} and some background structure $\gamma$, can be seen as a function on an appropriate n-th  jet-space of $\Phi$. Using the manifold structure of $\Phi$ we can take directional derivatives of $S$ and indeed of $\mathcal{L}$. 

Taking the functional derivative of $S$ along some generic $\delta\varphi^I$, for which we employ the standard notation for directional derivative, $\delta\varphi[S]$, we can usefully organize the result (using the jet-bundle structure) as: 
\be\label{eq:funct_L}
\dd S(\delta \varphi)=\int \sum_{k=0}^n\frac{\pp\mathcal{L}}{\pp(\nabla_{(\mu_1}\cdots\nabla_{\mu_k)}\varphi^I)}\nabla_{(\mu_1}\cdots\nabla_{\mu_k)}\delta\varphi^I= \int\left( E_I\delta\varphi^I+ \d(\theta_I(\varphi)( \delta\varphi^I))\right)
\ee
Here the Euler-Lagrange density is given by: 
\be E_I=\sum_{k=0}^n(-1)^k\nabla_{\mu_1}\cdots\nabla_{\mu_k}\mathcal{L}_I^{\mu_1\cdots \mu_k}
\ee 
where the partial derivative functions are given by: 
\be \mathcal{L}_I:=\frac{\pp\mathcal{L}}{\pp\varphi^I},\,\text{for $k=0$, and} \quad \mathcal{L}_I^{\mu_1\cdots \mu_k}:=\frac{\pp\mathcal{L}}{\pp(\nabla_{(\mu_1}\cdots\nabla_{\mu_k)}\varphi^I)}.
\ee
The symplectic potential  density $\theta$ is a $(d-1)$ differential form in spacetime and a 1-form in field-space (it is a local linear functional on field-space vectors $\delta\varphi$). And we can express the exterior derivative of a  $d-1$ form as the divergence of the corresponding vector:\footnote{This is done using the Hodge star, but we will keep notation as simple as possible and denote the vector-density form of $\theta$ with the same letter.}
\be\label{eq:Noether}\theta^\mu_I(\varphi)(\delta\varphi^I):=\sum_{k=1}^n\sum_{\ell=1}^{k}(-1)^{\ell+1}\left(\nabla_{\mu_2}\cdots\nabla_{\mu_\ell}\mathcal{L}_I^{\mu\mu_2\cdots \mu_k}\right)\left(\nabla_{\mu_{\ell+1}}\cdots\nabla_{\mu_k}\delta\varphi^I\right)
\ee

The idea now is simple. Suppose the Lagrangian in \eqref{eq:funct_L} is invariant in the direction of some particular $\widehat{\delta\varphi}$, and that the Euler Lagrange equations hold (which we signal by employing the weak equality $\approx$). 
If we use the coordinate-free notation of $\theta$ as a $d-1$ form, we conclude that the other term of the integrand of \eqref{eq:funct_L} weakly vanishes: 
\be\label{Noether}\d (\theta_I(\varphi)( \widehat{\delta\varphi}{}^I))\approx0,\ee
where d is the spacetime exterior derivative. Thus by integrating over a spacetime region we can obtain the usual regional conservation laws for  $\theta_I(\varphi)( \widehat{\delta\varphi}{}^I)$, which is called \textit{the Noether current density} for $\widehat{\delta\varphi}{}^I$. This is the essence of Noether's theorems.

\subsection{The covariant symplectic form and the Hamiltonian}\label{sec:cov_Ham}

Working now in field-space, we can introduce a symplectic structure as follows. Using the exterior derivative on the symplectic potential density, which is a  field-space one-form $\theta_I^\mu$, we obtain a symplectic-form density, $\omega_{IJ}^\mu:=(\dd \theta)_{IJ}^\mu$, which is a $d$-1 form in spacetime and a closed 2-form in field space (since $\dd \dd\equiv 0$). Choosing a closed Cauchy surface $\Sigma$ (for now unbounded), we define: 
\be\label{eq:sym_form} \Omega:=\dd \Theta, \quad \omega_{IJ}=\dd(\theta)_{IJ};\quad \Theta({\delta \varphi})= \int_\Sigma \theta_I{\delta \varphi}^I,\quad 
 \Omega({\delta \varphi}_1, {\delta \varphi}_2)= \int_\Sigma \omega_{IJ}{\delta \varphi}_1^I{\delta \varphi}_2^J\ee
where we have only kept the internal index $I$, which serves to remind us that we are dealing with components, also in the (suppressed) spacetime labels $\mu$ and $x$.

If the symplectic potential $\Theta$ is also  gauge-invariant on-shell (which is always the case if $\pp \Sigma=\emptyset$, (cf. \cite[eq. 3.22]{Lee:1990nz}) we can apply  the Cartan Magic formula 
 to obtain:
\be \bb L_{{\delta \varphi}}\Theta\approx 0=(\dd \fI_{{\delta \varphi}}+ \fI_{{\delta \varphi}}\dd)\Theta
\ee
and therefore we obtain $
\fI_{{\delta \varphi}}\Omega\approx-\dd (\Theta({\delta \varphi})), 
$ so that  the Hamiltonian generator for the flow ${\delta \varphi}$ is simply given on-shell by  $\Theta({\delta \varphi})$.\footnote{Using this formalism it is also easy to show, assuming the Poincar\'e lemma applies (so at least in finite-dimensions and for simply-connected regions) that if a closed 2-form $\Omega$  is left invariant by a flow $\delta\varphi$, then it can be used to obtain the symplectic generator of that flow, $f_{\delta \varphi}$. That is, $ \bb L_{{\delta \varphi}}\Omega=0=\dd(\fI_{{\delta \varphi}}\Omega)\Rightarrow \fI_{{\delta \varphi}}\Omega=\dd (f_{\delta \varphi})$. } Then one usually writes the Hamiltonian generator $H[\xi]$ of the gauge transformation $\xi$ as: 
\be\label{Ham_gen} H[\xi]=\int_\Sigma \theta_I\delta_\xi\varphi^I.
\ee 

Indeed, in the absence of boundaries of the Cauchy surface, we can go beyond the mere existence of a Hamiltonian generator: it is possible to show  that $\fI_{\widehat{\delta \varphi}}\Omega\approx 0$ for variations within the subset of field configurations which satisfy the equations of motion, in which case, the kernel $\widehat{\delta \varphi}{}$ is integrable in that subset (cf. \cite[eq. 3.27]{Lee:1990nz}).\footnote{Here is a simple proof: suppose a certain class of vector fields $\widehat{\delta \varphi}{}$ is such that $\fI_{\widehat{\delta \varphi}{}}\omega\equiv\omega_{IJ}\widehat{\delta \varphi}{}^J=0$. 
 Now, since $\omega$ is closed, using 
 the Cartan Magic formula (which is valid on general Banach manifolds), since $\dd\omega=0$: 
$$
\bb L_{\widehat{\delta \varphi}{}}\omega=(\dd \fI_{\widehat{\delta \varphi}{}}+ \fI_{\widehat{\delta \varphi}{}}\dd)\omega=0
$$
by which we can say that $\omega$ is $\widehat{\delta \varphi}{}$-invariant. Moreover, if we take the field-space commutator of $\widehat{\delta \varphi}{}_1, \widehat{\delta \varphi}{}_2$ (which we  denote by a ``double-struck'' square-bracket), i.e. $\lbr \widehat{\delta \varphi}{}_1, \widehat{\delta \varphi}{}_2\rbr=\bb L_{\widehat{\delta \varphi}{}_1}\widehat{\delta \varphi}{}_2$, and contract with $\omega$  we obtain:
$$ {\lbr \widehat{\delta \varphi}{}_1, \widehat{\delta \varphi}{}_2\rbr}^I\omega_{IJ}=(\bb L_{\widehat{\delta \varphi}{}_1}\widehat{\delta \varphi}{}_2^I)\omega_{IJ}=\bb L_{\widehat{\delta \varphi}{}_1}(\widehat{\delta \varphi}{}_2^I\omega_{IJ})-\widehat{\delta \varphi}{}_2^I \bb L_{\widehat{\delta \varphi}{}_1}\omega_{IJ}=0
$$
and therefore such vector fields in the kernel of $\omega$ form an integrable distribution by the Frobenius theorem.} More specifically, the on-shell integral manifold corresponding to the tangent vectors $\widehat{\delta \varphi}{}$ form `orbits' parametrized by gauge transformations.\footnote{Infinitesimally, each $\widehat{\delta \varphi}{}$ is of the form $\widehat{\delta \varphi}{}=\delta_\xi \varphi$ i.e. it is a result of an infinitesimal gauge-transformation $\xi$ on the field space $\Phi$.} It is then possible to constitute a finite equivalence relation along elements of the gauge-orbits and thus formally reduce the pre-symplectic to the symplectic formalism.

Moreover, note that applying a Lie derivative along $\widehat{\delta \varphi}{}$ to a gauge-invariant density $\theta$ and using $\fI_{\widehat{\delta \varphi}}\Omega\approx 0$, we obtain 
\be \bb L_{\widehat{\delta \varphi}{}}\theta=(\dd \fI_{\widehat{\delta \varphi}{}}+ \fI_{\widehat{\delta \varphi}{}}\dd)\theta\approx\dd ( \fI_{\widehat{\delta \varphi}{}}\theta)\approx0. \ee
Therefore, for bona-fide  gauge transformations---those which we can treat symplectically---the associated Hamiltonian  must be \textit{on-shell} field-independent. We can therefore weakly set the  Hamiltonian constraints corresponding to gauge transformations to zero: $H[\xi]\approx 0$. 

\section{Non-locality and Coulombic and radiative components }\label{sec:gf_Lor}

\subsection{Non-locality}
To fully comprehend the non-local content of the  equations of motion of gauge theories,  it is instructive to first examine an entirely local, extremely simple, example: that of a massless  Klein-Gordon field---$\varphi\in C^\infty(M, \bb C)$ such that $\pp^\mu\pp_\mu\varphi=0$. Given any initial values $(\varphi, n^\mu\pp_\mu \varphi)$  of configuration and configurational `velocity' on a Cauchy surface (or on a subset thereof),  we can find a corresponding solution of the Klein-Gordon equations in the domain of dependence of the region (by the Cauchy-Kowalevsky theorem). In this case, locality is a consequence of a single equation of motion, of the same order in spatial and time derivatives. 

But for gauge theories, some of  the  differential equations  of motion may contain only spatial and no time derivatives, signalling the occurrence of initial value constraints. This is the case with Yang-Mills theory. In other words, having chosen a Cauchy surface $\Sigma$, we are not free to choose any value we want for the doublet $(E^i_I, \rho_I)$ on $\Sigma$.

Let us take  the simple example of electromagnetism: the Gauss law $\pp^i E_i=\rho$ does not allow us to choose any value for the electric field on a given point $x\in \Sigma$ independently of the distribution of charges or of the values of $E$ in other points: there are Coulombic constraints that must be first satisfied. These constraints must be satisfied even before any  equation accounting for propagation in time is taken into account. Since the constraint is a spatial differential equation, resolving the constraint will require some amount of integration over a given region at a fixed time. 

To see that this requirement must be satisfied before any dynamics takes place, we can take an even simpler example: the electrostatic regime, in which case $E_i=\pp_i \phi$, where $\phi$ is the Coulombic potential. We then have to satisfy a Poisson equation, $\nabla^2\phi=\rho$, whose solution involves an integral of a Green's function over the entire $\Sigma$. 

But this non-locality need not be of an extreme sort! It won't allow faster-than-light-signalling for instance. And it doesn't necessarily involve knowledge of the charge distributions over the entire Cauchy-surface; one may exchange that completely global information for a more local one:  the electric flux on the boundary of a region surrounding our chosen point $x$. That is, given a point $x\in \Sigma$, and a closed region $R$  around $x$, knowing  both the charge distribution within $R$ and boundary conditions for $\phi$ on $\pp R$, we can establish the  electrostatic field at $x$. If one knows the `local electric flux' through $R$, $\pp_n\phi=E_n:=f$, where $n_i$ is the normal to $R$ within $\Sigma$,  then one knows the value of the electric field in $x$.\footnote{ While it is commonly said that if one satisfies the initial value constraint in an initial surface, the dynamical propagation equations will ensure  that it continues to be satisfied throughout time, this is only assured if there is never any flux at the boundary, or if $\Sigma$ is a closed surface. This is one of the consequences of the non-vanishing of the Hamiltonian generator of symmetries for bounded regions.}

Thus the static electric field within a region has \textit{regional} locality in the sense that it is determined by the distribution of charges in the region  \textit{and} boundary data. It is therefore a regional quantity, but not a strictly interior one. The boundary data represents the influence of the outside on the inside of the region. This provides a specification of degrees of freedom for each region which depends only on the interior of the region, and not on any boundary information. These are what we call ``radiative'' degrees of freedom.

 Any  new split of degrees of freedoom must be checked for consistency in at least two ways. First, is the split kinematically consistent? I.e. is it obeyed by symplectic conjugacy relations? And second: how independent or redundant are the different types of degrees of freedom, when taken regionally and globally? That is, when does knowledge of one sort of information determine the state of the other?  The first question will be taken up in section \ref{sec:rad_coul}, where we will prove consistency, and the second was taken in section \ref{sec:gluing}, where we showed that although for each region the radiative degrees of freedom is entirely functionally independent of the regional $\phi$,  the \textit{collection} of radiative degrees of freedom associated to a patchwork of mutually exclusive, jointly exhaustive regions  fully determines both the  radiative and the Coulombic degrees of freedom of the whole. There is thus a subtle manner in which the Coulombic components are superfluous only when the entire collection of regional interior data is available, but not before. This is consistent with previous work, \cite{GomesStudies}, arguing that gauge degrees of freedom are superfluous only from a global perspective, but not from a parochial one.
 
 \subsection{Coulombic and radiative symplectic pairs}\label{sec:rads}
To establish our claims, we must further examine the source of non-locality in the dynamics of Yang-Mills theory, and that is what we will do in this section, which is based on \cite[Sec. 3]{GomesRiello_new}. According to the above, at least in the standard variables,\footnote{Our conclusions are valid also using a holonomy gauge-invariant basis, e.g. with Wilson loops.} the source lies in the Gauss laws. Thus the first order of business is to segregate the Coulombic degrees of freedom from the remaining ones in the electric field. We will see that in a concrete sense the Coulombic ones are the carriers of the boundary information of the field.  

\subsubsection{Coulombic and radiative decomposition}\label{sec:rad_coul}
We  start with the standard decomposition of the electric field: 
\be E_i^I=\dot A^I_i-\D_iA_0^I.
\ee
Coulombic terms are usually described as those given by the gradient of the potential, i.e. $\D_iA_0^I$.  But $E^I_i$ can also contain Coulombic terms, i.e. terms of the form $\D_i \xi^I$, for some Lie-algebra-valued function $\xi\in C^\infty(\Sigma, \mathfrak{g})$.\footnote{Since we will assume all these functions are Lie-algebra valued, we will drop the index $I$ to unclutter notation for what follows.} Lucky for us,  there is a straightfoward way to extract the  strictly Coulombic component of $E_i$ by employing a generalized Helmholtz decomposition. The decomposition  spits out a unique covariant gradient of a Coulombic potential and another component, covariant-divergence-free, which we will call \textit{radiative}, and denote  as $E_i^\text{rad}$.

In other words,  any $E_i$ in the region $R$, \textit{without any restriction on boundary values}, can be uniquely decomposed into: 
\be\label{eq:dec_rad}  E_i=E_i^\text{rad}-\D_i\phi.
\ee
That is, both $E_i^\text{rad}$ and $\phi$ can be uniquely written as functionals of $E_i$ (see \cite[Sec. 3.2]{GomesRiello_new}) and, under a gauge-transformation, $E_i^\text{rad}$ transforms like the electric field and $\phi$ tranforms like $A_0$.\footnote{Note that if $\Lambda$ is the full Coulombic potential  component of $E$ then $\Lambda=\phi-A_0$, and this component transforms covariantly as well (i.e. as the rest of $E$).}  For our purposes here, the precise form of these functionals is secondary; what is important to know is that the  range of values $\phi$ can take are left  unconstrained off-shell (i.e. before we consider how they react to charged matter), while $E_i^\text{rad}$ obeys the following equations: 
\be \begin{dcases}
\D^i E^\text{rad}_i = 0 & \text{in } R,\\
n^iE^\text{rad}_i = 0 & \text{at }\pp R.
\end{dcases}
\label{eq:horizontalpert}
\ee
Although the radiative components are non-local within the region, they do \textit{not} require any further information from outside the region.\footnote{It is useful to distinguish `boundary conditions' from `boundary restrictions'. Boundary conditions such as the above are formal: they enforce no boundary restriction of the possible states. That is because, as mentioned in \eqref{eq:dec_rad}, any $E_i$ can be decomposed using such a $E^\text{rad}_i$. See the Appendix. \label{ftnt:rest}}

In the next section, we will present the symplectic decoupling of the radiative and Coulombic degrees of freedom.  Moreover, since the symplectic potential is intimately related to the Noether charges (see section \ref{sec:Noether}), by studying the decomposition of the symplectic potential we  will  verify that their Coulombic and  radiative parts correspond to the ``problematic'', gauge-variant Noether charges, and the gauge-invariant, ``non-problematic'' Noether charges, respectively.

\subsubsection{Coulombic and radiative symplectic potentials}

The total symplectic potential, introduced in section \ref{sec:cov_form} for standard vacuum Yang-Mills for a Cauchy surface $\Sigma$ without boundaries is given by (see the first term on the rhs of \eqref{eq:YM_theta}):
\be\Theta=\int_\Sigma E\wedge\dd A\label{eq:symp}\ee
where the phase space $\mathcal{P}$ is given in the basis $(A, E)$. In the presence of boundaries, i.e. for $R\subset\Sigma$ such that $\pp R\neq \emptyset$,  we can still translate the radiative-Coulombic decomposition of  phase space variables into a  decomposition of the symplectic potential  into a radiative and a Coulombic component, i.e. $\Theta=\Theta^\text{rad}+\Theta^\text{Coul}$. We should note that the decomposition requires field-space one forms projecting both  $\delta A, \delta E$,\footnote{Where we use an affine identification of $\mathrm{T}_\varphi\mathrm{T}\Phi$ with $\mathrm{T}\Phi$ (as in e.g. $\mathrm{T}_x\bb R^n\simeq \bb R^n$).} onto their radiative, i.e. non-Coulombic components, and   onto the complementary Coulombic components. 

The Coulombic symplectic potential is given by: 
\be\label{eq:symp_V}
\Theta^\text{Coul}=\oint_{\pp R} f_I\phi^I.
\ee
 $\Theta^\text{Coul}$ is a 1-form on phase-space, only acting as a one-form on one of the phase space directions: along disturbances in $A$ (as in \eqref{eq:symp}). So  in \eqref{eq:symp_V} $\phi$ is a field-space one-form and $\phi(\delta A)$ is the `Coulombic extraction' of the field-space vector  $\delta A$. Since a given perturbation such as $\delta A=\D \xi$ is pure-gauge, an extraction of the `pure Coulombic' part of $\delta A$ is an extraction of its pure-gauge component, i.e. $\phi(\delta A)=\xi$. $f$ is  the electric flux at the boundary, i.e. the normal component of the electric field there  (for the complete symplectic treatment of the decomposition,   see  \cite[Sec. 3.6]{GomesRiello_new}). Note that the dynamical relevance of the Coulombic terms appears only at the boundary $\pp R$:  the electric flux  is conjugate to the restriction of the pure gauge (i.e. pure Coulombic) part of $A$ to the boundary.

The radiative component of the symplectic potential  is given by:\footnote{Here I limit myself to the vacuum Yang-Mills case. For the full case, see \cite[Sec. 3.6]{GomesRiello_new} and the appendix here.}
\be\label{eq:symp_H}
\Theta^\text{rad}=\int_R \left(E^\text{rad}_{iI}\dd A_\text{rad}^{iI}\right)
\ee
where $\dd A_\text{rad}=\dd A-\phi$, with $\phi$ as above (i.e. $\dd A_\text{rad}$ only tracks variance in the physical, gauge-invariant parts of a field-space vector $\delta A$, again because $\delta A=\D \xi$ is symplectically orthogonal to $E^\text{rad}_{iI}$).
  
 We  find that the radiative component of the symplectic potential already has all the required properties for a formal symplectic treatment of gauge symmetries, as discussed in section \ref{sec:Noether}. Namely,  it is gauge-invariant $\bb L_{\widehat{\delta \varphi}{}}\Theta^\text{rad}=0$ and the corresponding Hamiltonian vanishes on-shell $\Theta^\text{rad}(\delta_\xi\varphi)\approx 0$. 
  
  The radiative Noether current can be seen as an `appropriately' gauge-fixed version of the Noether current: the gauge-fixing is  perturbative, covariant with respect to gauge transformations of the perturbed configurations,  and does not require boundary restrictions (see footnote \ref{ftnt:rest}). 
  
  In sum: the  pair constituting the regional radiative phase space   
   is parametrized by a bona-fide symplectic pair. The total symplectic potential $\Theta$ decomposes into a strictly interior, radiative, gauge-\textit{invariant} part, $\Theta^\text{rad}$,  and a boundary, gauge-\textit{variant} Coulombic part, $\Theta^\text{Coul}$. 
   
   The radiative degrees of freedom  exhaust the interior  degrees of freedom, but neither their constraints nor their observable content  depend on further boundary data;  i.e. gauge-invariant observables are formed on the interior of each region.\footnote{Note also that the radiative degrees of freedom possess the local gauge-covariance properties of the electric field. That is, one can form gauge-invariant objects by forming local scalars (e.g. through tracing the product with another locally covariant object).} 


\end{document}